# Comment on "Inclusion of the backaction term in the total optical force exerted upon Rayleigh particles in nonresonant structures"

Amir M. Jazayeri*

*Department of Electrical Engineering, Sharif University of Technology, Tehran 145888-9694, Iran*

**Abstract**– The conventional dipole approximation (CDA) assumes that the EM fields a small particle generates in the presence of its surrounding material bodies are equal to the EM fields a point-like dipole generates in the *absence* of the material bodies. The authors of [Phys. Rev. A 98, 013806 (2018)] investigate a modified dipole approximation (MDA), which assumes that the EM fields the particle generates are equal to the EM fields a point-like dipole generates in the *presence* of the material bodies. The authors interpret the approximate EM force under the MDA as the sum of four terms named 'generalized gradient force', 'generalized radiation pressure', 'generalized spin curl force', and 'new force term'. I show that such an interpretation is wrong and misleading: the generalized gradient force is not a gradient force; the generalized radiation pressure is not a radiation pressure; and the generalized spin curl force is not a spin curl force. In the case where the particle interacts with only one resonant EM mode of small enough linewidth, which is usually called 'self-induced back-action trapping' in the literature, I rectify some of the statements in their paper, and clear up some common misconceptions in the literature. In the case where the particle does not interact with any EM modes of small enough linewidth, I show that the numerical examples in their paper are inconclusive, and more importantly, the MDA is in principle inaccurate when the CDA is inaccurate. It should be noted that finding the approximate EM force under the MDA is computationally as hard as finding the exact EM force.

# I. BACKGROUND

## A. Conventional dipole approximation (CDA)

The exerted time-averaged EM force (which is usually shortened to EM force) on a particle can be written as the integral of the time-averaged Maxwell stress tensor over any surface enclosing only the particle [1]. The conventional dipole approximation (CDA) assumes that the EM fields a small non-magnetic particle generates in the presence of its surrounding material bodies are equal to the EM fields a linearly polarizable point-like electric dipole generates in free space (viz., in the absence of the material bodies) [2,3]. The $i$ component of the EM force under the CDA ($F_{CDA,i}$) reads $0.5\,\text{Re}(\vec{p}_{CDA} \cdot \partial \vec{E}_0^* / \partial i)$, where $\text{Re}(e^{-i\omega_L t} \vec{p}_{CDA})$ is the electric dipole moment, $\text{Re}(e^{-i\omega_L t} \vec{E}_0)$ is the incident electric field (viz., the electric field in the absence of the particle, but in the presence of its surrounding material bodies), and the derivative is evaluated at the position of the particle center ($\vec{r}_p$) [4,5]. The electric dipole moment phasor ($\vec{p}_{CDA}$), which is found self-consistently, can be written as $\alpha \vec{E}_0(\vec{r}_p)$, where $\alpha$ is a coefficient named 'electric polarizability' [2,3]. For a spherical particle of radius $R$ and relative permittivity $\varepsilon$, the polarizability reads $\alpha = \alpha_0 / [1 - i\alpha_0 k_0^3 / (6\pi\varepsilon_0)]$, where $\alpha_0$ denotes $4\pi\varepsilon_0 R^3 (\varepsilon - 1)/(\varepsilon + 2)$, and $k_L = 2\pi / \lambda_L = \omega_L / c$ is the wavenumber of the driving laser in free space [2,3]. It should be noted that Eq. (13) in [6], which describes $\alpha_0$, is incorrect. The $i$ component of $\vec{F}_{CDA}$ can be rewritten as the sum of $0.25\,\text{Re}(\alpha)\partial(\vec{E}_0^* \cdot \vec{E}_0)/\partial i$ and $0.5\,\text{Im}(\alpha)\,\text{Im}(\vec{E}_0^* \cdot \partial \vec{E}_0 / \partial i)$, which are the $i$ components of the gradient force ($\vec{F}_G$) and radiation pressure ($\vec{F}_R$), respectively [4]. Some authors have decomposed the radiation pressure ($\vec{F}_R$) into two terms, and coined 'spin curl force' [5]. However, it should be noted that 'spin' is a misnomer for EM fields because it does not have the properties of spin of electrons [7].

The gradient force comes from the dependence of the EM energy on the position of the particle while radiation pressure (and the spin curl force) comes from the initial momentum of the photons interacting with the particle. The photons interacting with the particle are either absorbed or scattered by the particle. Absorption leads to $\text{Im}(\varepsilon)$ in $\alpha_0$, while scattering leads to $-i\alpha_0 k_0^3/(6\pi\varepsilon_0)$ in the denominator of $\alpha$. For the sake of simplicity, and in agreement with [6], I ignore $\text{Im}(\varepsilon)$.

### B. Self-induced back-action trapping (SIBA)

Self-induced trapping, which was introduced in 2006 [8] and later called 'self-induced back-action trapping' (SIBA) [9-13], is defined as optical trapping of a particle by a resonator whose resonance frequency ($\omega_r$) as a function of the position of the particle ($\vec{r}_p$) meets the condition $A \triangleq \omega_r(\infty) - \omega_r(0) \gg \kappa$, where $\vec{r}_p = 0$ is defined as the trapping point, and $\kappa$ denotes the linewidth (full width at half maximum) of the spectral density of the energy stored by the resonator. It is evident that such a large $A$ necessitates detuning the angular frequency of the driving laser ($\omega_L$) from $\omega_r(\infty)$ by an amount $\Delta < 0$. In contrast to the statement made in Introduction of [6], SIBA in the sense that it has been theorized in [13] (viz., when the particle interacts only with an EM mode of a resonator) does not lead to an enhancement of the trapping force [14]. More precisely, the trapping force (normalized to $\alpha_0$) in the presence of a resonance frequency shift and detuning is smaller than or equal to the trapping force (normalized to $\alpha_0$) in the absence of any resonance frequency shift and detuning. However, by choosing an optimum value for $\Delta$, one can increase the width of the trapping potential [13], but it comes at the expense of a decrease in the depth of the trapping potential [14]. I will return to this point later. Also, I believe that SIBA is a misnomer, because $\omega_r$ is always sensitive to $\vec{r}_p$ (viz., $\partial \omega_r / \partial \vec{r}_p$ is always non-negligible) whenever the EM mode corresponding to $\omega_r$ contributes to the exerted force on the particle. Moreover, the exerted force on an object is always self-induced and thanks to the EM fields generated by the object even if the object is in free space (viz., even if no material bodies surround the object). Such an object in free space may be modeled by an electric dipole [5], a combination of an electric dipole and a magnetic dipole [4], a combination of two electric dipoles [15], etc.

### C. Failure of CDA

SIBA is one of the cases where the CDA fails. The important point about SIBA is that the CDA fails even if the particle is replaceable by a point-like dipole. The reason is that the *linewidth* of the EM mode is so small than the presence of the particle significantly changes the amplitude of the EM mode. The linewidth threshold for observing the failure of the CDA scales with $\alpha_0$.

When the particle does not interact with any EM modes of small enough linewidth, the CDA may still fail. If the particle has a large size or a large refractive index, the particle may not be replaceable by a point-like dipole [16]. Also, if the incident electric field ($E_0$) has strong spatial variations over the particle, the particle may not be replaceable by a point-like dipole [14].

### D. Modified dipole approximation (MDA)

Abbassi and Mehrany present a modified dipole approximation (MDA) in [6]. The MDA assumes that the EM fields the particle generates are equal to the EM fields a point-like electric dipole generates in the *presence* of its surrounding material bodies. Therefore, they write the electric field as the sum of the incident electric field ($\vec{E}_0$), the electric field the dipole generates in free space, and an electric field $\mathrm{Re}[e^{-i\omega_L t}\vec{\vec{G}}_s(\vec{r},\vec{r}_p)\vec{p}_{MDA}]$, where $\mathrm{Re}(e^{-i\omega_L t}\vec{p}_{MDA})$ is the electric dipole moment under the MDA, $\vec{\vec{G}}_s(\vec{r},\vec{r}_p)$ denotes the scattering Green's function of the material bodies surrounding the particle, $\vec{r}$ denotes the observation point, and $\vec{r}_p$ denotes the position of the particle center. For the sake of simplicity, and in agreement with the examples in [6], I assume that $\vec{E}_0$ is approximately parallel to the *x* axis, viz., $\vec{E}_0 \approx \hat{x} E_0$. Also, I ignore the non-diagonal elements of $\vec{\vec{G}}_s$ for $\vec{r} \approx \vec{r}_p$ (according to the reciprocity theorem, $\vec{\vec{G}}_s$ is symmetric for $\vec{r} \approx \vec{r}_p$). As a result, $\vec{p}_{MDA}$ is approximately parallel to the *x* axis, viz., $\vec{p}_{MDA} \approx \hat{x} p_{MDA}$. I will hereafter denote the *xx* element of $\vec{\vec{G}}_s(\vec{r},\vec{r}_p)$ for $\vec{r} \approx \vec{r}_p$ by $g(\vec{r})$.

The authors of [6] define an 'effective polarizability' $\alpha_{eff}$, and write $p_{MDA}$ as $\alpha_{eff} E_0$. The term 'effective polarizability' was coined in atomic physics [17,18], and later used in classical optics [15,19]. The effective polarizability, which is found self-consistently, reads $\alpha_0 / [1 - i\alpha_0 k_0^3 / (6\pi\varepsilon_0) - \alpha_0 g]$, where $g$ is evaluated at $\vec{r} = \vec{r}_p$. I will show that the use of $\alpha_{eff}$ is not only unnecessary, but it has also led to a serious misconception in [6].

## II. COMMENTS

### A. Physical meaning of MDA force terms

The $i$ component of the EM force under the MDA ($F_{MDA,i}$) can be written as the sum of $0.5\,\text{Re}(\alpha_{eff} E_0 \partial E_0^* / \partial i)$ and $0.5|\alpha_{eff} E_0|^2 \text{Re}(\partial g / \partial i)$, where the derivatives are evaluated at $\vec{r} = \vec{r}_p$. The authors of [6] interpret the latter as the $i$ component of a 'new force term' ($\vec{F}_N$). Also, they interpret the former as the sum of the $i$ components of a 'generalized gradient force' ($\vec{F}_{GG}$), a 'generalized radiation pressure' ($\vec{F}_{GR}$), and a 'generalized spin curl force' ($\vec{F}_{GS}$). The $i$ components of $\vec{F}_{GG}$ and $\vec{F}_{GR} + \vec{F}_{GS}$ read $0.25\,\text{Re}(\alpha_{eff})\partial|E_0|^2/\partial i$ and $0.5\,\text{Im}(\alpha_{eff})\text{Im}(E_0^* \partial E_0 / \partial i)$, respectively. It should be noted that the second line of Eq. (16) in [6], which describes $\vec{F}_{GS}$, is incorrect.

Interpretation of $\vec{F}_{MDA}$ as the sum of $\vec{F}_{GG}$, $\vec{F}_{GR}$, $\vec{F}_{GS}$, and $\vec{F}_N$ is a serious misconception. A gradient force must come from the dependence of the EM energy on the position of the particle while a radiation pressure (and a spin curl force) must come from the initial momentum of the photons interacting with the particle. Under the MDA, the electric field phasor is equal to the sum of the incident electric field phasor ($\vec{E}_0$), the electric field phasor the dipole generates in free space ($\vec{E}_1$), and the electric field phasor $\vec{E}_2 = \vec{G}_s \vec{p}_{MDA}$. The electric field phasor $\vec{E}_1$, which is singular at $\vec{r} = \vec{r}_p$, allows us to write $F_{MDA,i}$ as $0.5\,\text{Re}[\alpha(\vec{E}_0 + \vec{E}_2) \cdot \partial(\vec{E}_0^* + \vec{E}_2^*)/\partial i]$, where $\alpha$ denotes the conventional polarizability in the CDA. I rewrite $F_{MDA,i}$ as the sum of a gradient force $0.25\,\text{Re}(\alpha)\partial|\vec{E}_0 + \vec{E}_2|^2/\partial i$ and a radiation pressure $0.5\,\text{Im}(\alpha)\text{Im}[(\vec{E}_0^* + \vec{E}_2^*) \cdot \partial(\vec{E}_0 + \vec{E}_2)/\partial i]$. However, rewriting $F_{MDA,i}$ as the sum of $F_{GG,i}$, $F_{GR,i}$, $F_{GS,i}$, and $F_{N,i}$ is misleading. What the authors of [6] call 'new force term' ($\vec{F}_N$) is in fact a summation of a gradient force and a radiation pressure (and a spin curl force). More importantly, what they mistakenly call

'generalized gradient force' ($\vec{F}_{GG}$), 'generalized radiation pressure' ($\vec{F}_{GR}$), and 'generalized spin curl force' ($\vec{F}_{GS}$) are not a gradient force, a radiation pressure, and a spin curl force, respectively.

## B. Computational difficulty of MDA

From a computational viewpoint, finding $g$ at some observation points in the vicinity of the particle is as hard as finding the Maxwell stress tensor at those points unless there exists an analytical solution to $g$. In other words, finding the approximate EM force under the MDA ($\vec{F}_{MDA}$) is in principle as hard as finding the exact EM force.

## C. Applying MDA to SIBA

If the particle is replaceable by a point-like dipole, and interacts with only one resonant EM mode, $\alpha_0 g(\vec{r})$ in the MDA can be written as $iAu(\vec{r})u^*(\vec{r}_p)/(\kappa/2-i\Delta)$ in terms of the normalized electric field profile $u(\vec{r})$ of the EM mode. The definitions of $A$, $\kappa$, and $\Delta$ are the same as the definitions given in [6] and Section I of this Comment. Also, I assume that SIBA takes place, viz., $A \gg \kappa$.

Since SIBA takes place, the CDA fails, even though the particle is replaceable by a point-like dipole. The reason is that $\kappa$ is so small that the presence of the particle significantly changes the amplitude of the EM mode. The change in the amplitude of the EM mode can be simply derived by using the concept of energy [13] without invoking the MDA.

The authors of [6] highlight the difference between $\alpha_{eff}$ and $\alpha_0$ with an emphasis on $\Delta = 0$. They state that for $\Delta = 0$, $\alpha_{eff}$ is pure imaginary, and reads $i\kappa\alpha_0/(2A|u(\vec{r}_p)|^2)$. This statement means that for $\Delta = 0$, $\text{Im}(\alpha_{eff})$ is very large when $u(\vec{r}_p)$ approaches zero (viz., when

the particle approaches a node of the mode profile). This is incorrect because $\alpha_{eff}$ is pure real and equal to $\alpha_0$ when $u(\vec{r}_p)$ approaches zero, whether or not $\Delta$ is zero.

More importantly, from their statement, the implication is that $\text{Re}(\alpha_{eff})$ is non-negligible unless $\Delta$ is zero. However, $\text{Re}(\alpha_{eff})$ is in fact negligible in comparison with $\text{Im}(\alpha_{eff})$ if the conditions $-A \leq \Delta \ll 0$ and $\Delta \approx -A|u(\vec{r}_p)|^2$ are met. When these conditions are met, since $1-\alpha_0 \text{Re}(g)$ and $\alpha_0 \text{Im}(g)$ are both very small, one might mistakenly think that $\text{Re}(\alpha_{eff})$ is very large. However, it should be noted that when those conditions are met, $[\alpha_0 \text{Im}(g)]^2 / [1-\alpha_0 \text{Re}(g)]$ is approximately unity, and therefore, $\text{Re}(\alpha_{eff})$ is negligible in comparison with $\text{Im}(\alpha_{eff})$, which is very large.

They state in Introduction of their paper that SIBA in the sense that it has been theorized in [13] (viz., when the particle interacts only with an EM mode of a resonator) leads to an 'enhancement of the trapping force'. This statement, which is a common misconception, is incorrect. When the conditions $-A \leq \Delta \ll 0$ and $\Delta \approx -A|u(\vec{r}_p)|^2$ are met, what they call 'generalized gradient force' ($\vec{F}_{GG}$) is negligible, but what they call 'new force term' ($\vec{F}_N$) is non-negligible because a very large $|\alpha_{eff}|^2$ compensates for a very small $|E_0|^2$. The condition $\Delta \approx -A|u(\vec{r}_p)|^2$ means that the detuning $\Delta$ exactly compensates for (viz. is equal to) the resonance frequency shift $\omega_r(\vec{r}_p) - \omega_r(\infty)$ caused by the presence of the particle at $\vec{r}_p$. The detuning $\Delta \approx -A|u(\vec{r}_p)|^2$ is the detuning which provides the maximum achievable trapping force when the particle is at $\vec{r}_p$. However, the maximum achievable trapping force (normalized to $\alpha_0$) is equal to the trapping force (normalized to $\alpha_0$) in the absence of any resonance frequency shift and detuning. In other words, SIBA (viz., the existence of a large resonance frequency shift, and the necessity of a detuning) does *not* lead to an enhancement of the trapping force. I say 'normalized to $\alpha_0$' because the linewidth threshold for observing the failure of the CDA scales with $\alpha_0$.

### D. Applying MDA to nonresonant trapping schemes

The authors of [6] present two examples of the cases where the CDA is inaccurate. They emphasize that the particle in these examples does not interact with any EM modes of small enough linewidth. Also, they conclude that the exact EM force ($\vec{F}$) in such cases is approximately equal to what they call 'new force term' ($\vec{F}_N$).

The results of their first example are inconclusive for the simple reason that it lacks the calculation of $\vec{F}$. Their first example has been devised in a way that $\vec{F}_{MDA} - \vec{F}_{CDA}$ becomes comparable to $\vec{F}_{CDA}$. To this end, they consider a particle of a large diameter 220 nm and a large refractive index 2.5 at $\lambda_L$=1064 nm. However, the size of such a particle is larger than half the wavelength of the light inside it, and therefore, the missing calculation of $\vec{F}$ may show that the particle is not replaceable by a point-like dipole at all.

Also, it is noteworthy that their first example is not a practical trapping scheme because, in an attempt to make $\vec{F}_{MDA} - \vec{F}_{CDA}$ comparable to $\vec{F}_{CDA}$, they *decrease* the gradient force calculated within the CDA ($\vec{F}_G$) the by *not* allowing the structure surrounding the particle to see the laser light and contribute to the incident electric field ($E_0$). If there was no gap between the mirrors in their example (viz., if $\theta_0$ was zero), the structure would see the laser light and contribute to $E_0$. In such a case, not only would $\vec{F}_G$ be multiplied by a factor of 4, but also the unwanted radiation pressure would be zero.

The results of their second example are inconclusive for two reasons. First, it is true that $\vec{F}_{MDA}$ is a better approximation than $\vec{F}_{CDA}$, but their own numerical results show that $\vec{F} - \vec{F}_{MDA}$ is considerable when $\vec{F} - \vec{F}_{CDA}$ is considerable. They observe that what they call 'generalized gradient force' ($\vec{F}_{GG}$) is approximately equal to the gradient force calculated within the CDA ($\vec{F}_G$). Second, it is impossible to explain $\vec{F} - \vec{F}_{CDA}$ by what they call 'new force term' ($\vec{F}_N$) in the cases where the sign of $F_i - F_{CDA,i}$ depends on the size of the particle, because the sign of $F_{N,i}$ is independent of the size of the particle. One example is the trapping scheme proposed in [14], in which $F_{CDA,i}$ overestimates $F_i$ (viz., $F_{CDA,i} F_i > 0$ and $|F_{CDA,i}| > |F_i|$) with a large percent error for a very small particle whose diameter is ten nanometers, while $F_{CDA,i}$ underestimates $F_i$ (viz., $F_{CDA,i} F_i > 0$ and $|F_{CDA,i}| < |F_i|$) for a small particle whose diameter is a few tens of nanometers ($\lambda_L$=1550 nm, and the refractive index of the particles is 2 in [14]). Interestingly, it is also impossible to explain $\vec{F} - \vec{F}_{CDA}$ by $\vec{F}_{GG} - \vec{F}_G$ in such trapping schemes, because $\text{Re}(\alpha_{eff})$ cannot be positive and smaller than $\alpha_0$ for a particle of a certain size and refractive index, and larger than $\alpha_0$ for a larger particle of the same refractive index. In fact, the spatial variations of the incident electric field ($E_0$) over the particle in such trapping schemes are so strong that the particle is not replaceable by a point-like dipole, and the CDA and the MDA are both inaccurate.

Also, it is noteworthy that the second example in [6] is not a purely nonresonant trapping scheme. Rather, it is considered as a resonant trapping scheme by other authors [9]. The point that the resonant EM mode of the structure contributes to $\vec{F}$ is evidenced by the fact that the authors of [6] themselves consider a detuning between the angular frequency of the driving laser ($\omega_L$) and the bare resonance frequency of the structure (viz., the resonance frequency in the absence of the particle) in their second example.

When the CDA is inaccurate for a particle which does not interact with any EM modes of small enough linewidth, it seems impossible to *predict* whether the MDA is accurate (viz., whether the particle is replaceable by a point-like dipole). The reason is that the difference between $\vec{F}_{MDA}$ and $\vec{F}_{CDA}$ scales with $\alpha_0^2 |g|$ (for small $\alpha_0 |g|$). In other words, an increase in $\left|\vec{F}_{MDA} - \vec{F}_{CDA}\right|$ generally requires an increase in the size or refractive index of the particle, or a decrease in its distance from its surrounding structure. However, a particle of a large size or a large refractive index may not be replaceable by a point-like dipole [16]. Also, a particle close to its surrounding structure may not be replaceable by a point-like dipole either, because the EM fields close to its surrounding structure may have strong spatial variations over the particle [14].

\* jazayeri@ee.sharif.edu